\documentclass[
sigconf,
]{acmart}

\usepackage[utf8]{inputenc}
\usepackage{amsmath}
\usepackage{natbib}
\usepackage{graphicx}
\usepackage{tabularx}
\usepackage{caption}
\usepackage{subcaption}

\graphicspath{{figures/}}

\usepackage{times}
\usepackage{latexsym}
\usepackage{booktabs}
\usepackage{multicol}
\usepackage{multirow}
\usepackage{color}
\usepackage{url}

\usepackage{graphicx}

\newcommand{\LR}{\mathrm{LR}}
\newcommand{\FI}{\mathrm{FI}}
\newcommand{\FM}{\mathrm{FM}}
\newcommand{\FFM}{\mathrm{FFM}}
\newcommand{\IPNN}{\mathrm{IPNN}}
\newcommand{\ST}{\mathrm{ST}}
\newcommand{\CIN}{\mathrm{CIN}}
\newcommand{\softm}{\mathrm{softm}}
\newcommand{\AFM}{\mathrm{AFM}}
\newcommand{\EL}{\mathrm{EL}}
\newcommand{\ReLU}{\mathrm{ReLU}}
\newcommand{\FwFM}{\mathrm{FwFM}}
\newcommand{\AL}{\mathrm{AL}}
\newcommand{\SAMone}{\mathrm{SAM1}}
\newcommand{\SAMtwo}{\mathrm{SAM2}}
\newcommand{\SAMthree}{\mathrm{SAM3}}
\newcommand{\SAMAtwo}{\mathrm{SAM2_A}}
\newcommand{\SAMEtwo}{\mathrm{SAM2_E}}
\newcommand{\SAMAthree}{\mathrm{SAM3_A}}
\newcommand{\SAMEthree}{\mathrm{SAM3_E}}

\title{Looking at CTR Prediction Again: Is Attention All You Need? }
\author{Yuan Cheng, Yanbo Xue}
\thanks{Corresponding author: xueyanbo@kanzhun.com}
\affiliation{%
  \institution{Career Science Lab, BOSS Zhipin}
  \city{Beijing}
  \country{China}
}

\newtheorem{theorem}{Theorem}[section]
\newtheorem{proposition}[theorem]{Proposition}
\begin{abstract}

Click-through rate (CTR) prediction is a critical problem in web search, recommendation systems and online advertisement displaying. Learning good feature interactions is essential to reflect user's preferences to items. Many CTR prediction models based on deep learning have been proposed, but researchers usually only pay attention to whether state-of-the-art performance is achieved, and ignore whether the entire framework is reasonable. In this work, we use the discrete choice model in economics to redefine the CTR prediction problem, and propose a general neural network framework built on self-attention mechanism. It is found that most existing CTR prediction models align with our proposed general framework. We also examine the expressive power and model complexity of our proposed framework, along with potential extensions to some existing models. And finally we demonstrate and verify our insights through some experimental results on public datasets.

\end{abstract}



\copyrightyear{2021} 
\acmYear{2021} 
\setcopyright{acmcopyright}
\acmConference[SIGIR '21] {Proceedings of the 44th International ACM SIGIR Conference on Research and Development in Information Retrieval}{July 11--15, 2021}{Virtual Event, Canada.}
\acmBooktitle{Proceedings of the 44th International ACM SIGIR Conference on Research and Development in Information Retrieval (SIGIR '21), July 11--15, 2021, Virtual Event, Canada}
\acmPrice{15.00}
\acmISBN{978-1-4503-8037-9/21/07} 
\acmDOI{10.1145/3404835.3462936}

\begin{document}
\fancyhead{}
\maketitle

\section{Introduction}

With the booming of web 2.0, it is becoming more and more convenient for users to shop products, read news, and find jobs online. For service providers to attract and engage their users, they often rely on personalized recommendation systems to rank a small amount of items from a large amount of candidates. To achieve this goal, predicting user's behavior specifically via click-through rate (CTR) prediction becomes increasingly important. Therefore, effectively and accurately predicting CTR has attracted widespread attentions from both researchers and engineers.

From the perspective of a machine learning task, CTR prediction can be viewed as a binary classification problem. Classical machine learning models have played a very important role in the early adoption of CTR models, such as logistic regression (LR) models \cite{becker2007modeling, lr2007b, lr2010, lr2013}. Because linear models work under the strong assumption of linearity, a lot of and sometimes tedious feature engineering efforts are necessary to generate features that can be interacted linearly. To relax this constraint, a factorization machine (FM) model \cite{fm2010, fm2011, fm2012} was proposed to  automatically learn the second-order feature interactions. FMs and their extensions provide a popular solution to efficiently using second-order feature interaction, but they are still on the second-order level. For this reason, some deep neural networks (DNNs) are introduced to realize more powerful modeling ability to include high-order feature interactions. Among them, the factorization-supported neural network (FNN) \cite{fnn2016} is the first deep learning model that uses the embedding learned from FM to initialize DNNs, and then learns high-order feature interactions through multi-layer perceptrons (MLPs). 

Meanwhile, deep learning has successfully marched into many other application fields \cite{nature2015}, especially computer vision (CV) \cite{resnet2016} and natural language processing (NLP) \cite{bert2018}. Deep learning algorithms enable machines to perform better than humans in some specific tasks \cite{alphago2016}. Deep learning techniques have become the method of choice for working on the tasks of recommendation systems, but some researchers argue that the progress brought by deep learning is not clear \cite{rs2019a} and many deep learning models have not really surpassed traditional recommendation algorithms such as item-based collaborative filters \cite{rs2019a, rs2019b} and matrix factorizations \cite{rs2020}. Deep learning is usually branded as a black box due to the gap between its theoretical results and empirical evidences. For example, in terms of a recommendation system, DNNs usually involve implicit nonlinear transformations of input features through a hierarchical structure of neural networks. Finding a unified framework that can explain why it works (or why it does not) has become an important mission faced by many researchers. As yet another attempt, this paper aims to re-examine existing CTR prediction models from the perspectives of feature-interaction-based self-attention mechanism. 

Our goal for this work is to unify the existing CTR prediction models, and form a general framework using the attention mechanism. We  divide our framework into three types, which encompass most of the existing models. We use our proposed framework to extend the previous  models and analyze the CTR models from perspectives of theoretical and numerical results. From our research, we can classify almost all second-order feature interaction into the framework of the attention mechanism, therefore attention is indeed all you need for feature processing in CTR prediction. Our proposed framework has been validated on two public datasets. 

Four major contributions of our work are:
\begin{itemize}
\item  We use the discrete choice model to redefine the CTR prediction problem, and propose a general neural network framework with embedding layer, feature interaction, aggregate layer and space transform.
\item We propose a general form of feature interaction based on the self-attention mechanism, and it can encompass the feature processing functionalities of most existing CTR prediction models.
\item  We examine the expressive ability of the feature interaction operators in our framework and propose our model to extend the previous models. 
\item  Using two real-world CTR prediction datasets, we find our model can achieve extremely competitive performance against most existing CTR models.
\end{itemize}

The remainder of this paper is organized as follows. In \autoref{sec:related_work}, we surveyed existing models related to CTR prediction. Our proposed model is developed in \autoref{sec:model}, followed by a detailed analysis of its expressive power and complexity in \autoref{sec:math_analysis_sam}. Extensive experiments are conducted in \autoref{sec:experiments} to validate its performance. After discussing the implication of our work in \autoref{sec:discussion}, we conclude this paper in \autoref{sec:conclusions}.

\section{Related work}\label{sec:related_work}
Effective modeling of the feature interactions is the most important part in CTR prediction. Earlier attempts along this line include factorization machines and their extensions, such as higher-order FMs (HOFMs) \cite{hofm2016}, field-aware FMs (FFMs) \cite{fafm2016}, and field-weighted FMs (FwFMs) \cite{fwfms2018}. At the rise of deep learning models, deep neural networks have provided a structural way in characterizing more complex feature interactions \cite{nfm2017}. 

In addition to the depth, some researchers proposed to add width to the deep learning model. As such, Wide \& Deep model \cite{cheng2016wide} was proposed as a framework that combines a linear model (width) a DNN model (depth). Through joint training of the wide and deep parts, it can be better adapted to the tasks in recommendation system. Another example is the DeepCross model \cite{dc2016} for ads prediction, which shares the same designing philosophy as Wide \& Deep other than its introduction of residual network with MLPs. However, the linear model in the Wide \& Deep model still need feature engineering. To alleviate this, DeepFM model \cite{deepfm2017} was proposed to replace the linear model in Wide \& Deep with FMs. DeepFM shares the embedding between FMs and DNNs, which affects features of both low-order and high-order interactions to make it more effective.

At the same time, rather than leaving the modeling of high-order feature interactions entirely to DNNs, some researches are dedicated to constructing them in a more explicit way. For example, product-based neural network (PNN) \cite{pnn2016} was proposed to perform inner and outer product operations by embedding features before MLP is applied. It uses the second-order vector product to perform pairwise operations on the FM-embedded vector. The Deep and Cross Network (DCN) \cite{dcn2017} can automatically learn feature interactions on both sparse and dense input, which can effectively capture feature interaction without manual feature engineering and at a low computational cost. Similarly, in order to achieve automatic learning the explicit high-order feature interaction, eXtreme Deep Factorization Machine (XDeepFM) is proposed. In XDeepFM, a Compressed Interaction Network (CIN) structure is established to model low-level and high-level feature interactions at the vector-wise level explicitly. However, efforts spent on modeling high-order interactions might be easily dispersed since some researchers consider that the effect of higher than the second-order interactions on the performance is relatively small \cite{fb2019}. 

Thanks to the success of transformer model \cite{transformer2017} in NLP, the mechanism of self-attention has attracted some researchers in recommendation systems. To solve the problem that in FM model all feature interactions have the same weight, the Attentional Factorization Machine (AFM) model \cite{afm2017} was proposed, which uses a neural attention network to learn the importance of each feature interaction. Another work, known as AutoInit \cite{autoint2019}, was also inspired by the multi-headed self-attention mechanism in modeling complex dependencies. Base on a wide and deep structure, AutoInt can automatically learn the high-order interactions of input features through the multi-headed self-attention mechanism and provide a good explainability to the prediction results as well. 

All existing works, seemingly disconnected from each other, can somehow be brought under the same framework, which is the main contribution of our work to this community.

\section{Model}\label{sec:model}
\subsection{Problem Formulation}
For item $j \in Q$ and user $i \in P$,  $y_{i,j} \in \{0,1\}$ indicates whether the $i$-th user has engaged with the $j$-th item, with $Q$ and $P$ being the collections of items and users, respectively. In CTR prediction, engagement can be defined as clicking on an item. Our goal is to predict the probability of $p_i$ engaging with $q_j$. Obviously, this is a supervised binary classification problem. Each sample is composed of input of features $X = (X_{p_i}, X_{q_j})$ and output of a binary label $y_{i,j}$. The machine learning task is to estimate the probability for input $X$ as follows,
\begin{equation}
   \mathrm{Pr}\left(y_{i,j} = 1 | X_{p_i}, X_{q_j}\right) = F\left(X_{p_i}, X_{q_j}\right)
   \label{eq:ctr1}
\end{equation}
where $X_{p_i}$ is the feature of user $i$, and $X_{q_j}$ is the feature of item $j$.

\subsection{Discrete Choice Model}
CTR prediction problem corresponds to an individual's binary choice. We can use a discrete choice model (DCM) \cite{dcmbook} to describe this. DCM has found its wide range of applications in economics and other social science studies \cite{dcm2009}. 

The choice function of user $i$ belonging to $\Pi_i:U \rightarrow A$, where $U = \mathbb{R}$ is the utility space and $A$ is the users' choice sets $\{0: \textrm{not click}, 1: \textrm{ click}\}$. Let us define the utility obtained by user $i$ to choose item $j$ as follows,
\begin{equation}
u_{i,j} = H\left(X_{p_i},X_{q_j}\right)-\theta_{i, j} + k_i \epsilon_i
\label{eq:dcm}
\end{equation}
where $H(X_{p_i},X_{q_j})$ is the deterministic utility and $\theta_{i,j}$ is the expected utility, both indicating the $i$-th user choosing the $j$-th item. Here $\epsilon_i$ is a unit noise following a standard Gumbel distribution and $k_i$ is the noise level indicating uncertainty in the choice of user $i$. 

We can use a logit-based DCM to describe the user’s behavior. The probability of user $i$ selecting item $j$ can be expressed as,
\begin{equation}
w_{i,j} = \frac{1}{1+\exp\left(\frac{\theta_i-H(X_{p_i},X_{q_j})}{k_i}\right)}.
\label{eq:sigmoid}
\end{equation}

In the CTR prediction problem, features of the users and items are treated as a whole, \emph{i.e.}, $X = (X_{p_i}, X_{q_j})$, with which \autoref{eq:sigmoid} can be re-written as 
\begin{equation}
w_{i,j} = \sigma\left(M(X)\right),
\label{eq:G}
\end{equation}
where $M(X) = (H(X)-\theta_i)/k_i$ and $\sigma(x) = 1/(1 + \exp(-x))$. $M(X)$, as a nonlinear utility, can be defined as $M(X) = F_{\mathrm{NN}}(X)$ using a neural network structure as shown in  \autoref{fig:framework}.  Therefore, learning in a recommendation system is equivalent to obtaining the function $M(X)$.  

The binary cross-entropic loss can be obtained by maximum likelihood method, which is defined as follows,
\begin{equation}
\mathcal{L} = -\frac{1}{N}\sum_{i,j}\left[y_{i,j} \log w_{i,j} + (1 - y_{i,j}) \log(1 - w_{i,j})\right].
\label{eq:logloss}
\end{equation}
The above loss function is called log-loss, which is widely used in CTR prediction models.

\subsection{A General Neural Network Framework}\label{sec:general_framework}
\begin{figure}[th]
    \centering
    \includegraphics[width=\linewidth]{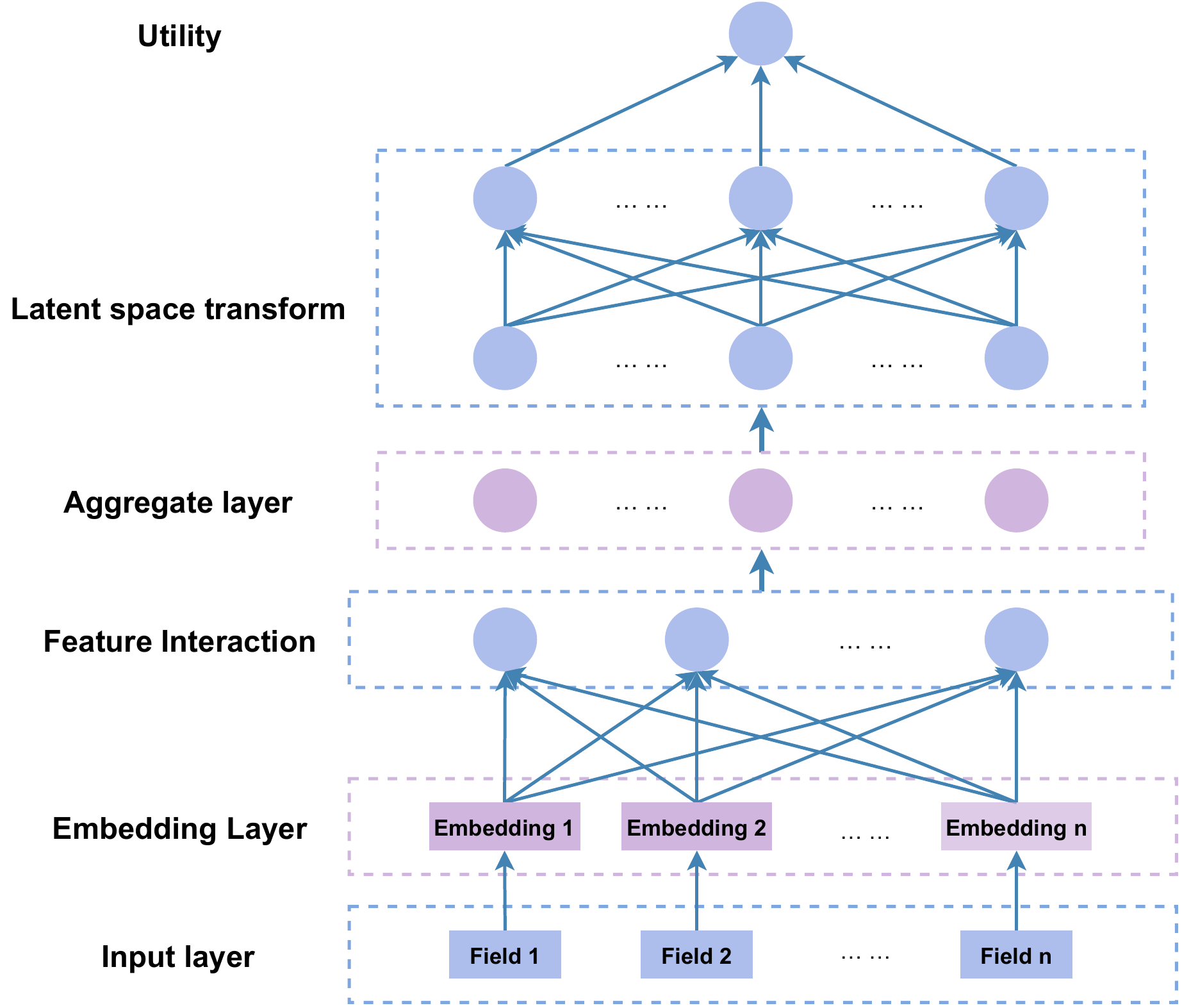}
    \caption{Overview of general framework of CTR prediction.}
    \label{fig:framework}
\end{figure}

Our proposed neural network framework is illustrated in \autoref{fig:framework}. For the sake of clarity, we only show main parts of the framework. The linear regression part as well as the skip connection similar to many previous models have been ignored.

\subsubsection{Embedding layer (EL)} In this work, only categorical features are considered, and numeric features can be converted into categorical data through discretization. Each feature can be expressed as a one-hot encoding. It is  assumed that the features have $n$ fields as $X = (x_1, x_2, ..., x_n)$.
The one-hot encoding $x_i$ can be converted into a vector in a latent space through embedding operation as follows
\begin{equation}
f_i = F_{\mathrm{emb}}(x_i) = W_{i}^{T} x_i
\label{eq:emb}
\end{equation}
where $W_i$ is the embedding matrix corresponding to the look-up table of the $i$-th field. In this work, the latent space is called utility space $\mathbb{R}^{d}$. 
After embedding operations, we can represent the categorical data $x_i$ as a vector $f_i \in \mathbb{R}^{d}$ 
in the $d$-dimensional utility space. Totally $n$ fields can be denoted as $f = [f_1, f_2, \cdots, f_n]$ and we denote $\{f_1, f_2, \cdots, f_n\}$ as $\mathcal{F}$.

\subsubsection{Feature interaction (FI)}
This part corresponds to the individual’s comprehensive measurement of the influence of different factors in the decision-making process. Due to that the relationship between the factors considered in the individual’s decision-making process is not independent \cite{fm2010}, FM has done a pioneering work in considering the second-order feature interactions.
  
The feature interaction layer is responsible for the second-order combination between features. The output is a $k$-dimensional vector. This layer is responsible for the second-order combination between features. Inspired by self-attention mechanism \cite{transformer2017}, a second-order operator of vector $v$ taking action on feature $f_i$ can be written as follows
\begin{equation}
 b_{S, U, v}(f_i) = S(f_i, v) \cdot U(f_i, v)
\label{eq:b}
\end{equation}
where $S(\cdot, \cdot)$ is a similarity function to measure the correlation degree between $f_i$ and $v$, and its value range is $[-\infty, \infty]$. And $U(\cdot, \cdot)$ is an utility function that indicates an individual utility induced by vector $v$. The utility function is a vector-valued function. When the dimension is 1, it is reduced to a scalar-valued function.

\autoref{eq:b} represents the utility vector obtained by the feature $f_i$ induced by vector $v$. The utility vector on $f_i$ induced by multiple vectors in $V$ can be expressed as
\begin{equation}
B_{S, U, V}(f_i) = \sum_{v \in V} b_{S, U, v}(f_i) = \sum_{v \in V}S(f_i, v) \cdot U(f_i, v)
\label{eq:B}
\end{equation}
where $S$ is the similarity function which can be viewed as weights for the outcomes in \autoref{eq:B}. In case that the weights are required to be positive, we can apply a softmax function. For convenience, we can denote $S_{s}$ as 
\begin{equation}
S_{s}(f_i, v) = \langle f_i, v\rangle_{\softm} = \frac{\exp\left\{S(f_i, v)\right\}}{\sum_{v \in V} \exp\left\{S(f_i, v)\right\}}.
\label{eq:softmax}
\end{equation}

The most common similarity function is the inner product operator $\langle\cdot,\cdot \rangle$. When the value of $S$ does not change, $S(f_i, v) = w$ becomes a constant-valued function, denoted as $w$. If $S(f_i, v) = 1$, we directly denote $S$ as $1$. The most common forms of utility function (score function) are linear function and constant-valued function. We set $1(v) = 1 $ as $1$, the linear function as $L$, and the identity function of $I(v) = v$  as $I$.

This part actually defines an attention mechanism between $f_i$ and $V$. If $V = \{f_1, f_2, \cdots, f_m\}$, the feature interaction reduces to self-attention effects. For simplicity, we denote $B_{S, U, V} = B_{S, U}$ as feature interaction via self attention in the rest of this work.

\subsubsection{Aggregation layer (AL)}
Feature interaction can process input features with $n$ fields $f = [f_1, f_2, \cdots, f_n]$ into utility vectors of $n$ fields $z = [z_1, z_2, \cdots, z_n]$, where $z_i = B_{S, U, V_i}(f_i) \in \mathbb{R}^{d}$. The role of the aggregation layer is to summarize the utility vectors of the $n$ fields into a utility vector. Common aggregation methods include concatenation and field combination, expressed as  
\begin{equation}
A_{C}(f) = \mathrm{vec}[f_1, f_2, \cdots, f_n]
\label{eq:concat}
\end{equation}
and
\begin{equation}
A_{L}(f) = \sum_{i = 1}^{n} w_i f_i
\label{eq:linear}
\end{equation}
respectively. Field combination (\autoref{eq:linear}) is a linear combination of the utility vectors in $n$ fields. In addition, we use $A_{\mathrm{mean}}(f) = \frac{1}{n} \sum_{i = 1}^{n} f_i$ and $A_{\mathrm{sum}}(f) = \sum_{i = 1 }^{n} f_i$ to denote the mean and sum of the $n$ fields.

\subsubsection{Space transformation (ST)}
After transformations of the feature interactions, the features have been converted from the original input space to the utility space $\mathbb{R}^d$. Assuming that the individual can transform in the utility space during the decision-making process, we use the structure of MLPs to define such conversion. After the input utility vector $z^{(0)}$ goes through a $k$-layer transformation, we can obtain
\begin{equation}
z^{(k)} = M^{(k)}(x)=L_{k}(a_{k}(L_{k-1}(a_{k-1}(\cdots(z^ {(0)})))))
\label{eq:mlp}
\end{equation}
where $L$ is a linear transformation, and $a$ is a non-linear activation function, and we use $M^{(0)}$ to represent $L$. In this study, unless stated otherwise, we set $a(x) = \ReLU(x)$ and $L(x) = W^T x + b$.

So far, we have developed the backbone module of our proposed framework, and there are other functional operators that also play an important role in existing CTR models, including regularization methods like layer normalization, batch normalization, dropout, and $L_2$ regularizer, and connection with network structure like skip connection $T_{F}(x) = F(x) + x$.

We use the framework established in \autoref{fig:framework} to decompose a CTR prediction model into
\begin{equation}
    M(X) = \ST \circ \AL \circ \FI \circ \EL(x)
    \label{eq:ctr}
\end{equation}
where $\EL$ corresponds to embedding layer, $\FI$ is the transformation of feature interaction, $\AL$ is the aggregation layer, and $\ST$ indicates the spatial transformation. 

\subsection{Feature Interaction in CTR Models}\label{sec:ctr_models}
In this work, we focus on second-order feature interactions, which is the most effective and widely used in CTR prediction models. Using the unified framework shown in \autoref{fig:framework} and specifically through feature interaction of $\FI = B_{S,U,V}$, we can reformulate the feature interaction layer of most existing CTR models as follows:

\subsubsection{Logistic Regression (LR)} LR model considers each feature independently, expressed as
$\phi_{\LR}(x; f) = \sum_{i = 1}^{n} x_i f_i$ where $f_i \in \mathbb{R}$. Therefore, for LR, feature interaction means $\FI_{\LR}(f_i) = f_i = B_{1, I, \{f_i\}}(f_i)$. Meanwhile, the similarity function and the utility function are reduced to 1 and $I$ respectively, and $f_i$ corresponds to $w_i$.

\subsubsection{Factorization Machine (FM) } FM enhances the linear regression model by incorporating the second-order feature interaction. FMs can learn the feature interaction by decomposing features into the inner product of two vectors as follows
$\phi_{\FM}(x; f) = \sum_{i < j}^{} x_i x_j \langle f_i, f_j \rangle$. And then we can find that the feature interaction in FM can be denoted as $FI_{FM}(f_i) = \sum_{j\neq i} \langle f_i, f_j \rangle \cdot 1 = B_{\langle, \rangle, 1, \bar{f_i}}(f_i)$. The similarity function is inner operator, the utility function is reduced to $1$, and $V = \bar{f}_i$ = $\{f_1, f_2, \cdots, f_n\} - \{f_i\}$. 

\subsubsection{Field-aware Factorization Machine (FFM)} 
Each feature belongs to a field. The features of one domain often interact with features of other different fields. By obtaining the embedding vector for $n-1$ fields of each feature, we can only use a vector $v_{i, F(j)}$ to interact with features $j$ in the field $F(j)$ as follows
$\phi_{\FFM}(x; f, F) = \sum_{i <j}^{} x_i x_j \langle f_{i, F (j)}, f_{j, F(i)} \rangle$ where $F(i)$ indicates the field to which the feature $i$ belongs. We can find that 
$$FI_{\FFM}(f_{i,F(j)}) = \sum_{j \neq i} \langle f_{i, F(j)}, f_{j, F(i)} \rangle \cdot 1 = B_{\langle, \rangle, 1, \bar{f}{_{i, F(j)}}} (f_{i,F(j)})$$.

\subsubsection {Field-weighted Factorization Machine (FwFM)} FwFM is an improvement to FFM to model the different feature interactions between different fields in a much more efficient way expressed as
$\phi_{\FwFM}(x; f, r) = \sum_{i< j}^{} x_i x_j \langle e_{i}, e_{j} \rangle w_{F(i), F(j)}$. And we can obtain $\FI_{\FwFM}(f_i) = \sum_{j\neq i} \langle f_i, f_j \rangle \cdot w_{F(i), F(j)} = B_{\langle, \rangle, w_{F(i), F(j)}, \bar{f_i}}(f_i) $, where the utility function becomes $w_{F(i), F(j)}$. 

\subsubsection{Product-based Neural Network (PNN)}
PNN is able to capture the second-order feature interactions through the product layer, which can take the form of Inner Product-based Neural Network (IPNN) or Outer Product-based Neural Network (OPNN). Since OPNN involves the operation of the aggregation layer, we focus on  IPNN, $\phi_{\IPNN}(x; f) = \sum_{i = 1}^{n} \sum_{j = 1}^{n} x_i x_j \langle f_{i}, f_{j} \rangle \langle \theta_{i}, \theta_{j} \rangle$, and we can find that $\FI_{\IPNN}(f_i) = \sum_{i, j} \langle f_i, f_j \rangle \cdot \langle \theta_i, \theta_j \rangle = B_{\langle, \rangle, \langle \theta_i, \theta_j \rangle, \bar{f_i}}(f_i) $ where the utility function is $\langle \theta_i, \theta_j \rangle$.

\subsubsection{Deep \& Cross Network (DCN)} 
DCN introduces a novel cross network (CN) \cite{xdeepfm2018} that is more efficient in learning certain bounded-degree feature interactions, which is defined as
$\phi_{\mathrm{CN}}(f) = w f$, \emph{i.e.}, $\FI_{\mathrm{CN}}(f_i) = w f_i = B_{1, w, f_i}(f_i)$ where utility function is $w$.

\subsubsection{DeepFM} 
As discussed previously, DeepFM combines the power of $\FM$  and MLPs into a new neural network architecture. Here we focus on the deep component which is the same as the Wide \& Deep model. This part was called implicit feature interaction through MLP in previous research. Using our framework, the feature interaction part is the same as LR, $\FI_{\mathrm{MLP}}(f_i) = f_i = B_{1, I, \{f_i\}}(f_i)$, and the implicit feature interaction is realized by the aggregation layer and the space transformation layer.

\subsubsection{XDeepFM}
The neurons in each layer of compressed interaction network (CIN) in XDeepFM are derived from the hidden layer of the previous layer and the original feature vectors. The second-order interaction part in CIN can be expressed as
$\phi_{\CIN}(f) = \sum_{i,j} p_{i,j} \langle A_{L_i}(f), A_{L_j}(f) \rangle$  where $A_{L_i }$ and $A_{L_j}$ are the field-wise aggregation operators and the feature interaction for the feature $f_i$ is
$\FI_{\CIN}(f_i) = \sum_{f_j} \langle f_i, f_j \rangle \cdot w_{i, j} = B_{\langle, \rangle, w_{i,j}, f}(f_i)$

\subsubsection{Attentional Factorization machine (AFM)}
AFM has one extra layer of attention-based pooling than FM. The function of the layer is to generate a weight matrix $a_{i,j}$ through the attention mechanism. The second-order interaction of AFM can be expressed as
$\phi_{\AFM}(x) = \sum_{i,j} a_{i, j} \langle f_i, f_j \rangle_{P}x_i x_j$. 
Here $a_{i,j}=e^{a_{i,j}'}/\sum_{i,j} e^{a_{i,j}'}$ and $a_{i, j}' = h^T \ReLU (W(f_i \odot v_j) x_i x_j + b)$. Therefore we can see that $\FI_{\AFM}(f_i) = \sum_{f_j \in \bar{f}_i} a_{i, j} \langle f_i, f_j \rangle_{P} \cdot 1 = B_{a_{i, j} \langle, \rangle_{P}, 1, \bar{f}_i}(f_i)$.

\subsubsection{AutoInt}
AutoInt can automatically learn the high-order interactions of the input features through multi-headed self-attention mechanism, expressed as
$\FI_{\mathrm{AutoInt}}(f_i) = \sum_{f_j \in f} \langle Q f_i, K f_j \rangle_{\softm} \cdot V f_i = B_{\langle Q\cdot, K \cdot \rangle _{\softm}, V, f}(f_i)$ with $\langle \cdot, \cdot \rangle_{\softm}$ being the softmax function defined in \autoref{eq:softmax}. 

\subsection{Self-Attention Feature Interaction}
Feature interaction is the key to the CTR prediction problem. Our work mainly focuses on second-order features interaction
\begin{equation}
    \FI(f_i) = \sum_{f_j \in V} \FI(f_i, f_j) = \sum_{f_j \in V} S(f_i, f_j) U(f_i, f_j)
    \label{eq:fi}
\end{equation}
where $f_i = \EL(x_i)$. $S(\cdot,\cdot)$ and $U(\cdot, \cdot)$ are defined similarly as in \autoref{eq:b}.

 \begin{table}[]
    \small
    \centering
    \caption{Unifying CTR models under one framework.}
    \label{tab:overview}
    \begin{tabularx}{1 \linewidth}{l  r  r  r  r  r r r}
        \toprule
        Model  & Input  &  $S$               & $U$    & $V$           & AL    & ST    \\
        \midrule
        LR     & $f_i$  & $1$                &  $I$   &  $\{f_i\}$    & $A_{\mathrm{sum}}$ &  /         \\ 
        FMs    & $f_i$  & $\langle, \rangle$ &  $1$   &  $\bar{f}_i$  & $A_{\mathrm{sum}}$ &  /    \\
        FFMs   & $f_{i,F(j)}$  & $\langle, \rangle$ &  $1$   &  $\bar{f}_{i,F(j)}$   & $A_{\mathrm{sum}}$ &  /    \\
       FwFMs   & $f_i$  & $\langle, \rangle$ &  $w_{F(i), F(j)}$  & $\bar{f}_i$  & $A_{\mathrm{sum}}$ &  /   \\
       IPNN    & $f_i$  & $\langle, \rangle$ &  $\langle \theta_i, \theta_j \rangle$  &  $\mathcal{F}$  & $A_C$   &  $M^{(k)}$\\ 
       DCN     & $f_i$  & $1$      & $w$   &  $\{f_i\}$   & $A_C$   &  $M^{(0)}$      \\
       DeepFM  & $f_i$  & $\langle, \rangle$ &  $1$   & $\bar{f}_i$ & $A_C$   &  $M^{(0)}$       \\
       XDeepFM & $f_i$  & $\langle, \rangle$   &  $w_{i,j}$   & $\bar{f}_i$ & $A_C$   &  $M^{(0)}$        \\
       AFM     & $f_i$  & $\langle, P \cdot \rangle$ &  $a_{i, j}$  &  $\bar{f}_i$  & $A_{\mathrm{sum}}$ &  /      \\
       AutoInt & $f_i$  & $\langle Q \cdot, P \cdot \rangle_{s}$ &  $V$   &  $\mathcal{F}$  & $A_C$ &  $M^{(0)}$       \\
         \bottomrule
    \end{tabularx}
\end{table}

We have defined a general neural network framework based on self-attention mechanism. As summarized in \autoref{tab:overview}, most CTR prediction models can be unified under this framework. Further more, models in \autoref{tab:overview} can be divided into three types:
\begin{itemize}
\item {\bf Type 1}: $\FI = B_{1, w, \{f_i\}} = w_i f_i$. In this case, the second-order feature interactions degenerate to first-order ones. Models like LR, DCN, and the wide component in Wide \& Deep and DeepFM belong to this type.

\item {\bf Type 2}: $\FI = B_{\langle, \rangle, w_{i, j}, \mathcal{F}}$. It is the FM model and its extensions, including FM, FFM, FwFM, IPNN, XDeepFM, and AFM. The characteristic of this type is that the similarity functions are all inner product operations $S(f_i, v) = \langle f_i, v \rangle$, and the utility function is a linear function with two variables in the form of $U(f_i, v) = w_ {i,j}(v)$ where $w_{i,j}(v) \in \mathbb{R}$.

\item {\bf Type 3}: $\FI = B_{{\langle Q\cdot, K \cdot \rangle _{s}, V, \mathcal{F}}}(f_i)$. This type uses self-attention mechanism in the transformer model, which contains AutoInt model. This type of model uses a similarity function as $S(f_i, v) = \langle Q f_i, K v \rangle$, and its utility function is a vector-valued function with one variable as $U(v) = V v$, where $v \in \mathbb{R}^{d}$ and $f_i \in \mathbb{R}^{d}$.

\end{itemize}

\subsection{Extension to CTR Models}
We can see that the most existing models can be divided into the above three types of $\FI$. As mentioned earlier, when self-attention is used, $B_{S,U,V}$ is simplified as $B_{S,U}$, we name such models as SAM, which means self-attention model. With SAM, a simple extension to these three types of models can be made by
\begin{equation}
    b_{\mathrm{SAM}}(f_i, f_j) = S(f_i, f_j) U(f_i, f_j),
\label{eq:sam_all}
\end{equation}
where $U(\cdot, \cdot)$ is a vector-valued function depending on $f_i$ and $f_j$.  In this work,  $U(\cdot, \cdot)$ takes one the two following forms, 
\begin{equation}
    U(f_i,f_j) = W_{i,j}
    \label{eq:sam1_u}
\end{equation}
and
\begin{equation}
    U(f_i,f_j) = f_i \odot f_j,
    \label{eq:sam2_u}
\end{equation}
where $W_{i,j} \in \mathbb{R}^{d}$ are trainable parameters, and $\odot$ indicates element-wise product of two vectors. When \autoref{eq:sam1_u} is used in SAM model, we call this kind of model $\mathrm{SAM_A}$, which means SAM with All trainable weights. When using \autoref{eq:sam2_u} in SAM, we obtain the model called $\mathrm{SAM_E}$, \emph{i.e.}, SAM by Element-wise product. Based on the general framework we proposed, we can further extend these three types of $\FI$.

\subsubsection{$\SAMone$} $\FI_{\SAMone} = B_{\langle, \rangle, 1}$. The form of $\FI$ in $\SAMone$ and $\LR$ model is exactly the same, except for its embedding dimension of $f$ changing to $d$. Then, we have
\begin{equation}
    \FI_{\SAMone}(f_i) = f_i 
    \label{eq:sam1_fi}
\end{equation}
with which we can obtain $\SAMone$ as follows,
\begin{equation}
    \SAMone(f) = M^{(0)} \circ  A_C \circ \FI_{\SAMone}(f),
    \label{eq:sam1}
\end{equation}
where $f = [f_1, f_2, \cdots, f_n]$, $A_C$ is the concatenation aggregate layer defined in \autoref{eq:concat}, and  $M^{(0)}$ is a linear transformation defined in \autoref{eq:mlp}. 

\subsubsection{$\SAMtwo$} $\FI_{\SAMtwo} = b_{\langle, \rangle, U_{i, j}}$. We can extend FM models to the following two forms,
\begin{equation}
    \FI_{\SAMAtwo}(f_i, f_j) = \langle f_i, f_j \rangle W_{i, j} 
    \label{eq:sama2_fi}
\end{equation}
and 
\begin{equation}
    \FI_{\SAMEtwo}(f_i, f_j) = \langle f_i, f_j \rangle f_i \odot f_i,
    \label{eq:same2_fi}
\end{equation}
with which, we can obtain $\SAMAtwo$ and $\SAMEtwo$ as follows,
\begin{equation}
    \SAMAtwo(f) = M^{(0)} \circ  A_C \circ \FI_{\SAMAtwo}(f)
    \label{eq:sama2}
\end{equation}
and
\begin{equation}
    \SAMEtwo(f) = M^{(0)} \circ  A_C \circ \FI_{\SAMEtwo}(f),
    \label{eq:same2}
\end{equation}
where, $\FI_{\SAMAtwo}(f) = [\FI_{\SAMAtwo}(f_i, f_j)]_{i,j} \in \mathbb{R}^{n \times n \times d}$ and 

\noindent $\FI_{\SAMEtwo}(f) = [\FI_{\SAMEtwo}(f_i, f_j)]_{i,j} \in \mathbb{R}^{n \times n \times d}$.

\subsubsection{$\SAMthree$} $\FI_{\SAMthree} = B_{{\langle Q\cdot, K \cdot \rangle _{\softm}, V}}(f_i)$. This type is closely related to self-attention mechanism in the transformer model. This type of model uses a similarity function of $S(f_i, f_j) = \langle  f_i, K f_j \rangle$ where two linear transformation are combined in the inner product, and we extend the original utility function of $U(f_j) = V f_j$ to $U(f_i,f_j) = W_{i,j}$ and $U(f_i, f_j) = f_i \odot f_j$, and then we can obtain
\begin{equation}
    \FI_{\SAMAthree}(f_i, f_j) = \langle  f_i, K f_j \rangle W_{i, j} 
    \label{eq:sama3_fi}
\end{equation}
and 
\begin{equation}
    \FI_{\SAMEthree}(f_i, f_j) = \langle  f_i, K f_j \rangle f_i \odot f_j.
    \label{eq:same3_fi}
\end{equation}

Inspired by the network structure of AutoInt \cite{autoint2019}, we propose two variants of $\SAMthree$ as follows
\begin{equation}
    \SAMAthree(f) = M \circ  A_L \circ (\FI_{\SAMAthree}^{(L)} + Q^{(L)}) \cdots  (\FI_{\SAMAthree}^{(1)} + Q^{(1)})(f) 
    \label{eq:sama3}
\end{equation}
and
\begin{equation}
    \SAMEthree(f) = M \circ  A_L \circ (\FI_{\SAMEthree}^{(L)} + Q^{(L)}) \cdots  (\FI_{\SAMEthree}^{(1)} + Q^{(1)})(f) 
    \label{eq:same3}
\end{equation}
where $L$ is the number of $\mathrm{SAM}$ layers, $Q$ is a linear mapping, and $A_L$ is a field combination aggregation. Without claimed explicitly, $L = 1$ and $M = M^{(0)}$  in this work.

\section{Mathematical Analysis of SAM}\label{sec:math_analysis_sam}

SAM has four parts as shown in \autoref{eq:ctr}. $\EL$ is embedding layer, $\FI$ is the transformation of feature interaction, $\AL$ is the aggregation layer, and $\ST$ indicates the spatial transformation. We denote the set of all the models satisfying the form in \autoref{eq:ctr} as $\mathcal{M}$.

\subsection{Expressive Power}
\begin{definition}\label{def1}
[Expressive power $\succeq_{\mathcal{M}}$] 
$\forall M_1 \in \mathcal{M}$ when trainable parameters in $M_1$ are determined, $\exists M_2 \in \mathcal{M}$ with certain parameters in $M_2$ and $\ST_2$ such that $\ST _2 \circ \AL_2 \circ \FI_2 \circ \EL_2 = \ST_1 \circ \AL_1 \circ \FI_1 \circ \EL_1$, then we can say that the expressive power of $M_2$ is higher than that of $M_1$, which is denoted as $M_2 \succeq_{\mathcal{M}} M_1$. 
\end{definition}

\begin{definition}\label{def2}
[Expressive power $=_{\mathcal{M}}$] 
$\forall M_1 \in \mathcal{M}$ and $M_2 \in \mathcal{M}$, if $M_1 \succeq_\mathcal{M} M_2$ and $M_2 \succeq_{\mathcal{M}} M_1$, it can be considered that the expressive power of $M_1$ is equal to that of $M_2$, which can be denoted as $\FI_1 =_{\mathcal{M}} \FI_2$. 
\end{definition}

Using Definitions \ref{def1} and \ref{def2}, we make three propositions:
\begin{proposition}
\label{p:1}
 $\SAMone =_{\mathcal{M}} \LR$.
\end{proposition}

\begin{proposition}
\label{p:2}
 $\SAMAtwo \succeq_{\mathcal{M}} \FM \succeq_{\mathcal{M}} \LR$. 
\end{proposition}

\begin{proposition}
\label{p:4}
 $\SAMAthree \succeq_{\mathcal{M}} \SAMAtwo$.
\end{proposition}

The above three propositions are easy to check and the proofs are thus omitted here. It is noted that the $\ST$ in SAMs is a linear transformation. The idea behind the proof is that when EL, FI, LA and ST are all linear operators, the trainable parameters can be aggregated together and absorbed by the free parameters in the last layer. From these propositions, we can obtain
\begin{equation}
  \SAMAthree \succeq_{\mathcal{M}} \SAMAtwo \succeq_{\mathcal{M}} \FM \succeq_{\mathcal{M}} \SAMone =_{\mathcal{M}} \LR.
  \label{eq:math}
\end{equation}

We see that if the deep learning method can find the global minimum of the CTR prediction problem, its expressive power can fully reflect the performance of the model. Therefore, we deduce that the potential of $\SAMAthree$ and $\SAMAtwo$ model will be greater than that of $\FM$ and $\LR$.

\begin{table}[h]
 \small
    \centering
    \caption{Summary of SAM complexities}
    \label{tab:complexity}
    \begin{tabularx}{0.8 \linewidth}{l  r  r}
        \toprule
        Model  & Space $O(\cdot)$      &  Time $O(\cdot)$ \\
        \midrule
        $\LR$     &  $n$       & $n$     \\ 
        $\SAMone$   &  $2d n$    & $dn$    \\
        \midrule
        $\FM$    & $n + d n$      &  $dn$      \\
        $\SAMAtwo$  & $2d n^2 + d n$ &  $2 dn^2$      \\
        $\SAMEtwo$  & $d n^2 + d n$  &  $2 dn^2$     \\
        \midrule
       AutoInt & $3L d^2 + 2 d n$        & $L(3 d^2 n + 2 d n^2) + d n$ \\
       $\SAMAthree$   & $L(d^2 + d n^2) + 2d n$ & $L(  d^2 n + 2 d n^2) + d n$ \\
       $\SAMEthree$   & $L d^2  + 2d n$     & $L(  d^2 n + 2 d n^2) + d n$ \\
         \bottomrule
    \end{tabularx}
\end{table}

\subsection{Model Complexity}

We analyze the space complexity and time complexity of $\SAMone$, $\SAMtwo$ and $\SAMthree$ models in terms of the four operators in \autoref{eq:ctr}. In $\mathrm{SAM}$, $n$ is the number of feature fields, $d$ is the embedding vector dimension and $L$ is the number of layers in $\SAMthree$. For the space complexity, we ignore the bias term in the linear transformation. $\EL$ is a shared component which contains $dn$ parameters. $A_C$ has no parameters and calculation overhead.  $\ST$ is a linear transformation, which has $dn$ parameters and the amount of computation is $O(dn)$ for $\SAMone$ and $\SAMthree$. And for $\SAMtwo$, $\ST$  needs to be calculated $O(dn^2)$ times with $dn^2$ parameters. 

The main difference between these three models lies in $\FI$. In $\SAMone$, $\FI$ has no extra space and time cost. In $\SAMtwo$, we need $n^2 d$ parameters for the weight vectors in $\SAMAtwo$ and no more space for $\SAMEtwo$. And the time cost is $O(2 n^2 d)$ for $\SAMtwo$. As for $\SAMthree$, for each layer, the linear transform spends $d^2$ parameters and extra $n^2 d$ for the weights in $\SAMAthree$. The time overhead of SAM3  mainly depends on the linear transformation $O(d^2 n)$ and the computation on attention $O(2 d n^2)$ for each layer.

Based on these analysis, we can get the model complexity results as shown in \autoref{tab:complexity}. The time and space complexities of the $\SAMone$ model are $d$ times those of LR, the $\SAMtwo$ model is about $n$ times that of FM, and the complexity of $\SAMthree$ and AutoInt is very close. Considering that both $d$ and $n$ are relatively small, our $\textrm{SAM}$ model has a certain computational efficiency.

\section{Experiments}\label{sec:experiments}
\subsection{Experiment Setup}
 \begin{table}[h]
    \small
    \centering
    \caption{Statistics of the datasets.}
    \label{tab:datasets}
    \begin{tabularx}{0.7 \linewidth}{l r  r  r}
        \toprule
        Dataset  &  \# Samples  & \# Categories & \# Fields\\
        \midrule
        Criteo   & 45,840,617 &  $1,086,810$      &  39    \\ 
        Avazu    & 40,428,967 & $2,018,012$ &  22    \\
        \bottomrule
    \end{tabularx}
\end{table}
\subsubsection{Datasets} In this section, we will conduct experiments to determine the performance of our model compared to other models. We randomly divide the dataset into three parts: 80\% for training, another 10\% for cross validation, and the remaining 10\% for testing. Table \ref{tab:datasets} summarizes the  statistics of the two following public datasets we have used in our experiments:
\begin{enumerate}
\item \textbf{Criteo}\footnote{http://labs.criteo.com/2014/02/kaggle-display-advertising-challenge-dataset/}: It includes one week of display advertising data, which can be used to estimate the CTR of advertising by CriteoLab, and it is also widely used in many research papers. The data contains the click records of 45 million users, which contains 13 numerical feature fields and 26 categorical feature fields. The numerical feature is discretized by the function $\mathrm{discrete}(x) = \lfloor2\times \log(x)\rfloor$ if $x>2$ and $\mathrm{int}(x-2)$ otherwise. 
\item \textbf{Avazu}\footnote{https://www.kaggle.com/c/avazu-ctr-prediction/data}: This is the data provided by Avazu to predict whether a mobile ad will be clicked. It contains 40 million users' 10 days of click log with 23 categorical feature fields. We remove the field of sample id which is not helpful to CTR prediction. 
\end{enumerate} 

\subsubsection{Evaluation Metrics} In the experiment, we use two evaluation indicators: AUC (Area Under ROC) and log-loss (cross entropy; \autoref{eq:logloss}). AUC is the area under the ROC curve which is a widely used metric for evaluating CTR prediction. AUC is not sensitive to classification threshold and a larger value means a better result. Log-loss as the loss function in CTR prediction, is a widely used metric in binary classification, which can measure the distance between two distributions a smaller value indicates better performance. 
 
\subsubsection{Baseline Models}  We have benchmarked our proposed $\mathrm{SAM}$ model against eight existing CTR models (LR, FM, FNN, PNN,  DeepFM, XDeepFM, AFM and AutoInit as described in \autoref{sec:ctr_models}) as well as an original transformer encoder with one layer and one head, and two higher-order models (AFM \cite{afm2017} and HOFM \cite{hofm2016}). For all deep learning models, unless explicitly specified, the depth of hidden layers is set to 3, the number of hidden layer neurons is set to 32, and all the activation functions are set as $\ReLU$. In terms of initialization, we initialize embedding vectors by Xavier's uniform distribution method \cite{xavier2010}. For regularization of all models, we use $L_2$ regularizer to prevent overfitting. Through performance comparisons on different validation sets, we choose to use $\lambda = 10^{-5}$. In addition, the dropout rate is set to 0.5 by default for some classic models which needs to use or not used otherwise.

\subsection{Performance Comparison}
\begin{table}[tp]
  \small
  \centering
  \caption{Overall performance on the datasets.}
  \label{tab:performance_comparison}
    \begin{tabularx}{0.8 \linewidth}{l l l l l }
   \toprule
    \multirow{2}{*}{Model}&
    \multicolumn{2}{c}{Criteo} & \multicolumn{2}{c}{Avazu} \cr \cline{2-5}
    &AUC&log-loss&AUC&log-loss\cr
    \midrule
        LR      &0.7949   &0.4555 &0.7584&0.3921 \\

        $\FM$      &0.8078   &0.4443 &\underline{0.7858}&\underline{0.3777} \\
        $\FFM$     &0.8077  &0.4438 &0.7742&0.3829 \\
        $\FwFM$    &{0.8089}&{0.4427} &0.7778&0.3810 \\

        $\IPNN$    &\underline{0.8107}&\underline{0.4408} &{0.7818}&{0.3791} \\
        DCN     &0.8074&0.4439   &0.7798&0.3800 \\
        DeepFM  &0.8030&0.4487   &0.7798&0.3799 \\
        XDeepFM &0.8104&0.4414  &0.7809&0.3798 \\

        $\AFM$     &0.8067&0.4448   &0.7775&0.3812 \\
        AutoInt    &{0.8106}&{0.4411}&{0.7834}&{0.3780} \\
       
        AFN     &{0.8097} &{0.4421} &{0.7809} &{0.3791} \\
        HOFM    &0.7993&0.4523  &0.7737&0.3837 \\
        Transformer&0.7942  &0.4566 &0.7693&0.3866 \\
        \midrule
        $\SAMone$    &0.7925&0.4572   &0.7720&0.3848 \\
        $\SAMEtwo$  &\bf{0.8115}&\bf{0.4404} &\bf{0.7891} &\bf{0.3755} \\
        $\SAMAtwo$  &0.8098&0.4420 &0.7885&0.3756  \\
        $\SAMEthree$  &0.8071&0.4451 & {0.7805}& {0.3821} \\
        $\SAMAthree$  &{0.8098}&{0.4420} & 0.7796 & {0.3805} \\
    \bottomrule
    \end{tabularx}
\end{table}

All models are implemented using neural network structures from \emph{PyTorch} \cite{pytorch2017}. The models are trained with \emph{Adam} optimization algorithm \cite{adam2015} (learning rate is set as $0.001$). For all models, the embedding size is set to 16, and the batch size is set to 1024. We conduct all the experiments with 8 GTX 2080Ti GPUs in a cluster setup. 

The results of the numerical experiments are summarized in \autoref{tab:performance_comparison}. The scores are obtained by 10 different runs for each category. The highest value across different models is shown in bold and the highest performance obtained by baseline is underlined. We have verified the statistical significance in our results with $p$-value $< 0.05$. We compared three proposed models, $\SAMone$, $\SAMtwo$, and $\SAMthree$, with 12 CTR prediction models as well as the transformer encoder in a simple structure of a single-layer encoder with one head. It can be found that our proposed $\SAMEtwo$ model performs the best on both Criteo and Avazu datasets. The second-order interaction models IPNN and FM also perform competitively on Criteo datasets and Avazu datasets respectively, and are even better than XDeepFM based on higher-order interactions in our experiments. Therefore, to a certain extent, it consolidates the fact that many CTR prediction problems mainly rely on the second-order feature interaction. The performance improvement brought by higher-order interaction such as XDeepFM, Transformer and HOFM under the existing framework may not be significant. It’s worth noting that AutoInt performs reasonably well on both datasets, which even rivals the popular Transformer model. This can be explained by the fact that, although layer normalization can reduce the bias shift, it has also induced correlations among features that a shallow model is unable to resolve. This also explains why our proposed single-layered $\SAMthree$ model does not perform well in general.

It can be found that the relationship we obtained in \autoref{eq:math} is not completely consistent with the results of numerical experiments. For example, the performance of $\SAMone$ in the Criteo dataset is slightly worse than that of $\LR$, but much higher than that of $\LR$ in the Avazu dataset. The performance of $\SAMAtwo$ is better than that of $\SAMone$ and $\FM$ models, and $\SAMAthree$ in the Avazu dataset is inferior to $\SAMAtwo$. From our experimental results, we can find that the models with over-parameters would have potential to get better performance.

\begin{figure}[th]
    \centering
    \begin{subfigure}{0.5\textwidth}
    \centering
    \includegraphics[width=\linewidth]{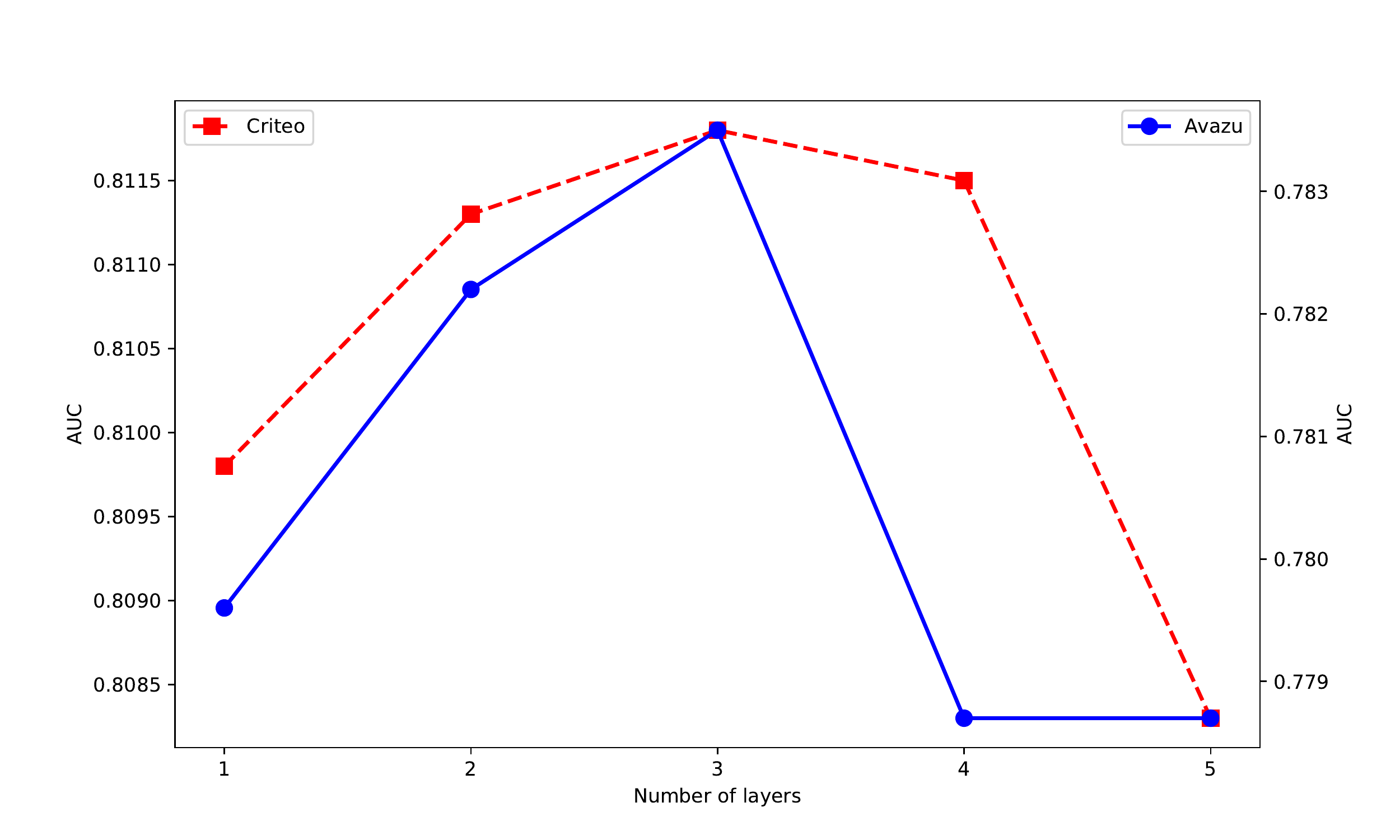}
    \caption{AUC}
    \end{subfigure}
    \begin{subfigure}{0.5\textwidth}
    \centering
    \includegraphics[width=\linewidth]{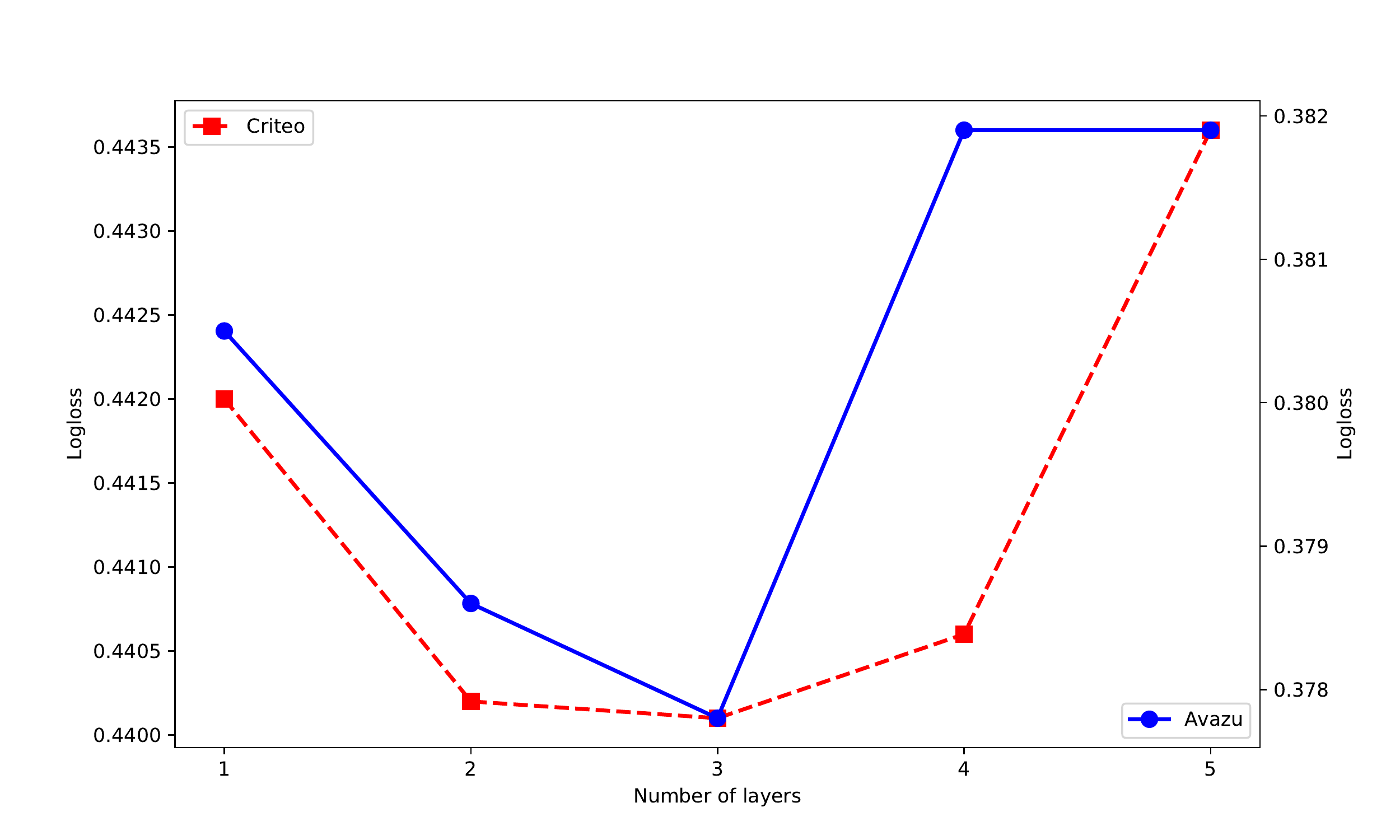}
    \caption{log-loss}
    \end{subfigure}
    \caption{Performance as a function of the number of layers in $\SAMAthree$.}
    \label{fig:auc}
\end{figure}

Since all weights are trainable, it is not surprising to observe that $\SAMAthree$ performed better than $\SAMEthree$, as evidenced by the last two rows of \autoref{tab:performance_comparison}. As part of an ablation study of $\SAMAthree$, we discussed the relationship between the number of layers and its performance. As shown in \autoref{fig:auc}, on both datasets, the performances of $\SAMAthree$ are consistent with the change of the number of layers. When the number of layer is 3, $\SAMAthree$ reaches its best performance. At this time, the AUC on the Criteo data set is 0.8118 and log-loss is 0.4401. It is 
slightly higher than the previous best result from $\SAMEtwo$. Better results are also obtained for the Avazu dataset, with an AUC of 0.7835 and a log-loss of 0.3778. This study provides us insights that for models such as $\SAMthree$, multiple layers of self-attention structure can improve the performance, but the excessively high-order feature interaction formed by too many layers will reduce the effect of the model.

\section{Discussions}\label{sec:discussion}
There is no doubt that over the last two decades, deep learning models have been very successful in the fields of CV and  NLP, which also make them a fundamental building block of feature extractions in recommendation systems. However, in industrial applications, both their working mechanisms and explanabilities are still being challenged from time to time \cite{rs2019a, rs2019b}, and sometimes even being \emph{outperformed} by classical machine learning methods like tree-based models \cite{rec2020}.

A recommendation system is completely different from the CV and NLP tasks. The main objectives in CV and NLP systems are mimicking the perceptual abilities of human beings, and recommendation systems is to understand the fundamental mechanisms in human's decision-making behavior. Its well known that as a high-level human cognitive functionality, human behavior is to difficult to model due to human's bounded rationality \cite{bounded2000}.

In this work, we are intended to provide a general framework to model human decision-making behaviors for CTR prediction problems. We proposed our extended models of $\mathrm{SAM}$s. We aimed at providing a general framework to further extend the CTR prediction model, rather than focusing on obtaining the state-of-the-art performance, and therefore, performance comparisons are not explored comprehensively in this work. It is often unstable to always use the powerful fitting ability of deep learning models to obtain a high performance even before fully understanding the human decision-making mechanism. Even if the results of state-of-the-art are obtained, it is blessed by the proper distribution of the dataset and laborious tunings of hyperparameters. Instead, we should pay more attention to human behaviors. When modeling with a deep learning framework, we will benefit more if we can open the black box and connect the network structure and its functionalities with the human decision-making process. As a preliminary attempt towards this direction, this work provides a unified framework and hopefully more researches can be extended on this basis.

\section{Conclusions}\label{sec:conclusions}
In this work, a general framework for CTR prediction is proposed, which corresponds to an individual decision-making process based on neural network model. We also attempt to study whether the attention mechanism is critical in the CTR prediction model. It is found that most CTR prediction models can be viewed as a general attention mechanism applied to the feature interaction. In this sense, the attention mechanism is of importance for CTR prediction models. 

In addition, we extend the existing CTR models based on our framework and propose three types of $\mathrm{SAM}$s, in which $\SAMone$ and $\SAMtwo$ models are extensions of $\LR$ and $\FM$ models, respectively, and $\SAMthree$ corresponds to the self-attention model in Transformer with original one-field embedding extended to pairwise-field embedding. According to the experimental results on the two datasets, although our extension can obtain quite competitive results, the $\SAMthree$ model has not demonstrated its significant advantages. We also perform a more in-depth analysis of the number of $\mathrm{SAM}$ layers in the $\SAMAthree$ model, and find that depth does not always lead to better performance. To a certain extent, this also shows that the CTR prediction problem is different from the NLP task, and the effect of high-order feature interactions cannot bring too much improvement. 

To conclude, we have established a unified framework for CTR prediction and a possible direction for future work should be on the combination of this framework to models that can help us to understand human decision-making behavior, {\em i.e}, agent-based model.

\section*{Acknowledgements}
We thank the anonymous reviewers for their valuable comments. We are grateful to our colleagues for their helpful discussions about CTR prediction problem and the self-attention mechanism.

\bibliographystyle{ACM-Reference-Format}
\bibliography{main}

\end{document}